# Experimental demonstration of the acoustic frequency conversions by temporal phononic crystals

Xujun Dong,[1] Yangtao Ye,[1] Bing Wang,[2] Chunyin Qiu,[1] Manzhu Ke,[1] and Zhengyou Liu[1,*]

[1]Key Laboratory of Artificial Micro- and Nano-structures of Ministry of Education and School of Physics and Technology, Wuhan University, Wuhan 430072, China

[2]Institute of Materials Research and Engineering, Agency for Science, Technology and Research (A*STAR), Singapore 117602

**Abstract:**

We report on the experimental investigation of the transmission spectra for acoustic waves through temporal phononic crystals (TPCs), which are designed structures with periodically time-varying density and bulk modulus. In our experiment, the TPCs are instanced as the media that swap between air and carbon dioxide repetitively, in between the sound source and the detector. The transmission spectra for monochromatic incident acoustic waves exhibit a series of equidistant peaks, manifesting novel up- and down-conversion effects while not employing material nonlinearity. These experimental results agree with the theoretical calculations based on the plane wave expansion in time domain.

**PACS numbers:** 43.20.Fn, 43.40.Le, 42.79.Nv

---

[*]To whom all correspondence should be addressed, zyliu@whu.edu.cn

In the past two decades there is a great deal of interest in photonic crystals [1] and phononic crystals [2], which have great advantage in molding the energy flow of light (or electromagnetic waves) and sound (or elastic waves) compared to conventional materials, and thus have potential applications in a variety of functional optical and acoustic devices. Both photonic crystals and phononic crystals are spatially periodic structures, which are respectively the periodic distributions of permittivity and/or permeability, and the periodic distributions of density and/or elastic modulus in space. The wave motions and their properties in spatially periodic structures are the main issues in the studies concerning photonic crystals or phononic crystals. Naturally, one may ask a question: how would waves propagate in a structure with periodically time-varying properties? Apparently, structures with periodically time-varying properties are analogous to photonic crystals or phononic crystals, except that the periodic variations of properties are in the time domain rather than in space. Recently, theoretical studies on structures with periodic time-varying electromagnetic properties [3–8], termed as temporal photonic crystals [4,5], had been presented. The dispersion relations of the electromagnetic waves in the temporal photonic crystals exhibit bands or band gaps in wave numbers, in contrast to the bands or band gaps in frequency for conventional photonic crystals. Actually, the earliest studies on electromagnetic waves propagating in structures with temporal properties can be traced back to a half century ago [9,10], and novel effects, such as frequency conversion or splitting had been predicted. It was proposed recently that these dynamic materials could be used to create optical isolation [3] and photonic Aharonov-Bohm effect [8]. Similarly, there were also theoretical studies on acoustic waves propagating in structures with periodically varying acoustic properties, i.e. temporal phononic crystals (TPCs), and frequency conversions were demonstrated [11].

While it is not very difficult to fabricate photonic crystals or phononic crystals, it is a great challenge to realize temporal photonic/phononic crystals in practice, because of their dynamical nature. To make material properties vary with time periodically as desired is in general not as easy as just to array them periodically in space [12–17]. That is why so far there are rare experimental studies on either temporal photonic crystals or phononic crystals. By applying external acoustic/electric field [12–16], the optical properties of acousto-optic/electro-optic materials can be changed. However, these changes are usually too weak (e.g., about 0.1% in permittivity [12]) to exhibit the unique features of temporal photonic crystals. For the same reason, it is also diffcult to realize a TPC with remarkable periodic changes in acoustic parameters. In this letter, we instance a TPC by swapping the medium between the source and detector from air to carbon dioxide and vice versa repeatedly, in order to achieve an adequate periodic time-variation in density and modulus. In this way, the changes in density or bulk modulus can reach as high as 10%, which allow the frequency converting effect to be demonstrated experimentally.

Our sample (Fig. 1(a)) consists of a hollow steel cylinder of diameter 40 cm and height 35 cm, divided into four identical chambers by four radial plates joined at the rotating axis. The cylinder is attached to an electric motor through the rotating axis. One chamber in the cylinder is filled with carbon dioxide ($CO_2$) and the other three are filled with air. To prevent the $CO_2$ from leakage, the front and back ends of the cylinder are sealed up with thin plastic films. A sound source (B&K Type 4206) is placed close to the front end of a chamber, which launches sound signals. In the opposite direction across the chamber, a detector (B&K Type 4187) is placed close to the back end of the same chamber, which receives the sound propagating through the sample. The collected signals are then analyzed by a multi-analyzer system (B&K Type 3560B). Both the sound source and detector are fixed in positions, and when the cylinder rotates, air and $CO_2$ swap periodically in the space between the source and the detector, which mimics a TPC with time-varying density and bulk modulus.

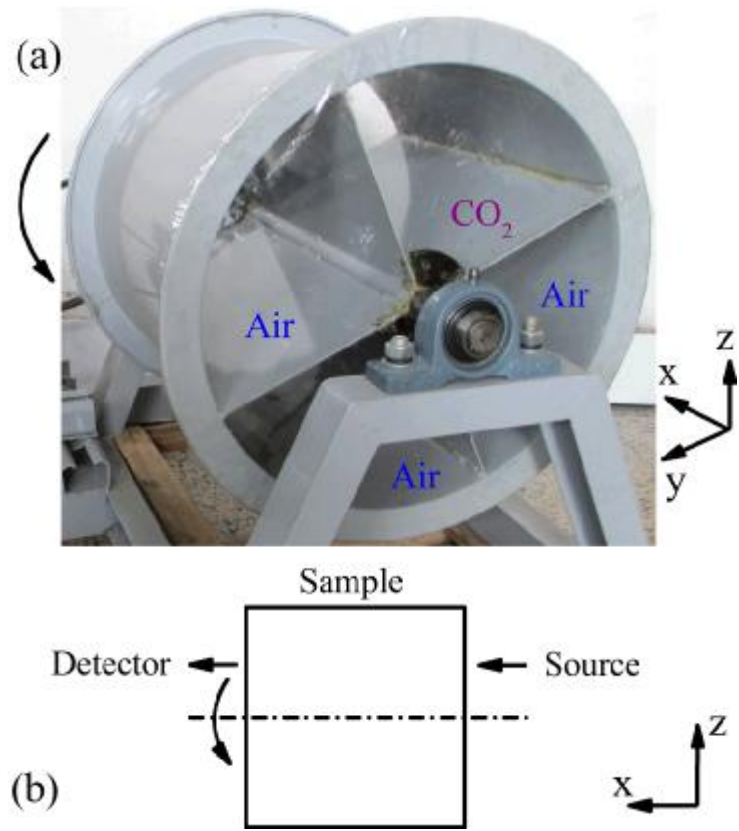

FIG. 1. Set-up of the experiment. (a) Photograph of the TPC sample: a rotating hollow cylinder with four chambers filled by air or carbon dioxide ($CO_2$). The front and back ends of the cylinder are sealed up with thin plastic films. The diameter and height of the cylinder are 40 cm and 35 cm, respectively. (b) Schematic illustration of the measurements performed.

Denoting the rotation period of the system with $T$, the time-variations of the density and bulk modulus of the TPC can be depicted in Fig. 2. It is convenient to introduce a concept of *time-filling-fraction*, to characterize the TPC, which is defined as $F = t_1/T$, with $t_1$ being the duration of the $CO_2$ in between the source and the detector in each period. Obviously, time-filling-fraction is an analogue of filling fraction used for characterizing spatial phononic crystals. In our experiment, $F = 1/4$ since only one chamber is filled with $CO_2$ while the other three are filled with air. Besides, the modulation-strengths of mass density ($\Delta r = (r_a - r_b)/r_b$, with $a$ for $CO_2$ and $b$ for air) and bulk modulus ($\Delta B = (B_a - B_b)/B_b$) for this sample are 53% and 10%, respectively.

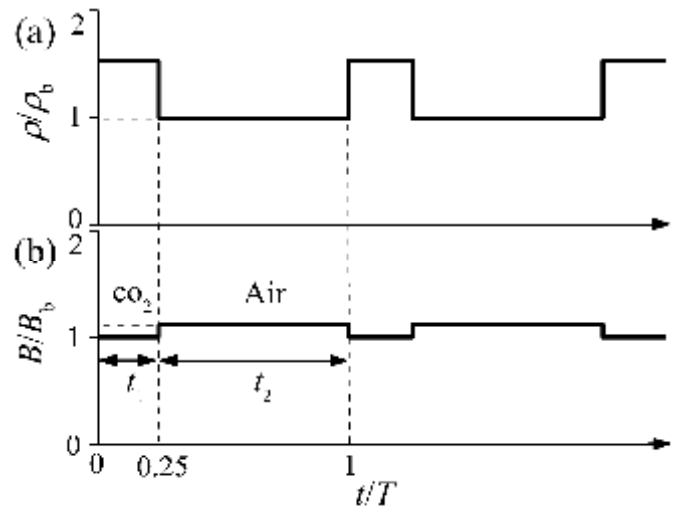

FIG. 2. Acoustic properties of the TPC varying with time $t$, (a) the density and (b) the bulk modulus. The TPC is instanced as the repeated swapping between air (with duration $t_1$) and $CO_2$ (with duration $t_2$), which have different densities and bulk moduli.

In the first measurement, the cylinder is set to rotate at a frequency of $\Omega = 9.3\,\text{Hz}$, which corresponds to a modulation period $T \approx 0.11\,\text{s}$, and a continuous sound of frequency $w_0 = 3.0\,\text{kHz}$ is launched into the sample. The measured transmission spectrum is shown in Fig. 3(a) with solid lines. As expected, the transmission spectrum consists of multiple equidistant peaks (solid lines in Fig. 3(a)), among which the fundamental peak locates at the frequency $w_0 = 3.0\,\text{kHz}$, i.e., the frequency of the incident waves. The other peaks or the *split* peaks are distributed at both sides of the fundamental peak, with the interval nearly equal to the modulation frequency $\Omega$. In the subsequent measurement, the rotating frequency of the cylinder is set to $\Omega = 12.4\,\text{Hz}$, and the frequency of the incident waves is tuned to $w_0 = 4.5\,\text{kHz}$, then similar transmission spectrum is obtained (the solid lines in Fig. 3(b)). This time, the fundamental peak moves to $w_0 = 4.5\,\text{kHz}$ and the peak interval increases to $\Omega = 12.4\,\text{Hz}$. In both measurements, the transmission spectra are normalized to the height of the corresponding fundamental peaks.

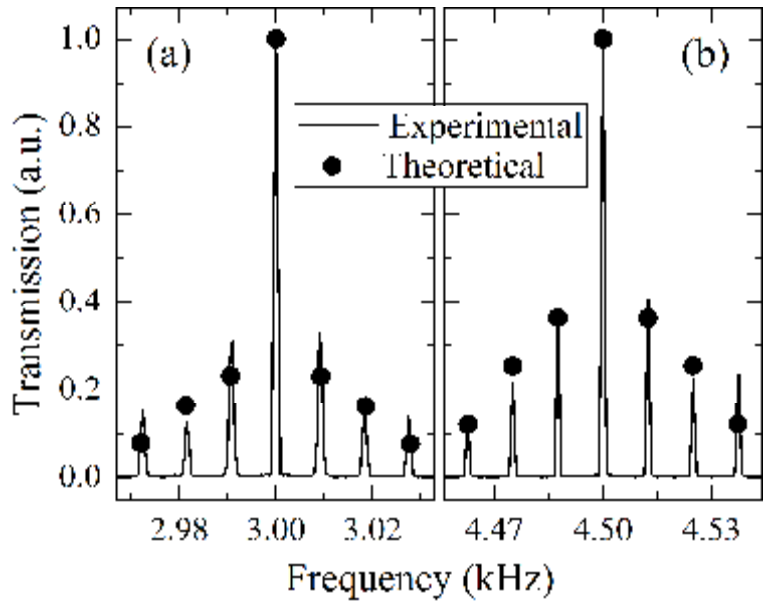

FIG. 3. Transmission spectra of experimental (solid lines) and theoretical (dots) results as a function of frequency. (a) The incident frequency $w_0 = 3.0$ kHz while the modulation frequency $\Omega = 9.3$ Hz. (b) The incident frequency $w_0 = 4.5$ kHz while the modulation frequency $\Omega = 12.4$ Hz.

Note that the temporal periodicity can be changed directly by tuning the speed of the motor. This flexibility is a remarkable inherent feature for the current TPC, as for conventional phononic crystals, the spatial periodicity cannot be changed after they have been fabricated.

To theoretically analyze the transmission of the sample, we need to obtain the band structure and the eigenmodes of the TPC in advance, which are necessary in the transmission calculation. Here the plane-wave expansion method in time domain [4] is employed to derive the band structure. Specifically, the density $r(t)$ and bulk modulus $B(t)$ are expanded as Fourier series $g(t) = \sum_n g_n e^{in\Omega t}$ with $g$ denoting $r$ or $B$, and the displacement field $u$, a Bloch function with time $t$, is expanded as $u(x,t) = \sum_m u_m e^{im\Omega t} e^{i(kx-wt)}$. Substituting these expansions into wave equation, $\frac{\partial^2 u(x,t)}{\partial x^2} = \frac{1}{B(t)} \frac{\partial}{\partial t} \left( r(t) \frac{\partial u(x,t)}{\partial t} \right)$, series of the following equation can be finally obtained:

$$\sum_{n=-\infty}^{\infty} \left\{ \sum_{l=-\infty}^{\infty} \left[ (w-n\Omega)(w-l\Omega) B_{m-l} r_{l-n} - d_{m,n} k^2 \right] \right\} u_n = 0, \qquad (1)$$

which give the band structure. Figure 4 shows the band structure of our TPC, and as can be seen, it is the wave number bands in the first Brillouin zone (BZ) of frequency, in contrast to the band structure of conventional phononic crystals, which is the frequency bands in the first BZ of wave number. For a given frequency $w_0$, e.g.

$w_0 = 3.0$ kHz as indicated by the dashed line in Fig. 4 when translated into the first BZ, there are a series of modes, each at one band with a specific wave number, e.g. $k = k_q$, with $q$ being the band number. For the mode with $w = w_0$ and $k = k_q$ at band $q$, the wave function, expressed as $u_{w_0 k_q}(x,t) = e^{ik_q x} \sum_m u_m(w_0, k_q) e^{-i(w_0 - m\Omega)t}$, can be obtained by substituting $w = w_0$ and $k = k_q$ into Eq. (1) and calculating the expansion coefficients $u_m(w_0, k_q)$. The wave function is essential in transmission calculation as depicted below.

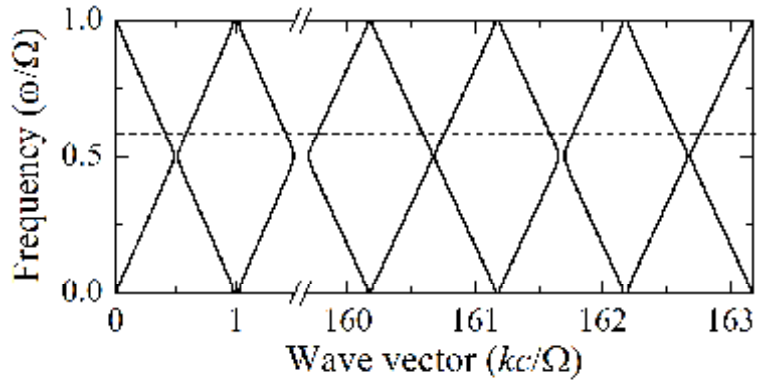

FIG. 4. Calculated band structure for the TPC, where the parameters are set as the same as in Fig 3(a). The dash line indicates the incident frequency ($w_0 = 3.0$ kHz) which is translated into the first Brillouin zone. Besides the first three bands, followed by a break, the 320$^{th}$ to the 326$^{th}$, the last seven bands are plotted, which dominantly account for the corresponding seven peaks in Fig 3(a).

For a finite slab of this TPC subject to a normal incident wave of frequency $w_0$ (see the experiment setup, Fig. 1(b)), the field inside the slab can be expressed as the linear combination of the eigenmodes with frequency $w_0$, at all bands propagating along both the positive and the negative directions, i.e.,

$$u(x,t) = \sum_{q=1}^{\infty}\left[a_q u_{w_0,k_q}(x,t) + b_q u_{w_0,-k_q}(x,t)\right] = \sum_{q=1}^{\infty}\left[a_q e^{ik_q x} + b_q e^{-ik_q x}\right]\sum_{m=-\infty}^{\infty} u_m(W_0,k_q)e^{-i(w_0-m\Omega)t}$$

where $k_q$ and $-k_q$ represent respectively the modes propagating from left to right and from right to left inside the slab. Since the incident wave can be expressed as $u^i(x,t) = u_0 e^{i(k_0 x - w_0 t)}$, and the reflected and transmitted waves can be expressed as

$$u^r(x,t) = \sum_{n=-\infty}^{\infty} u_n^r e^{i[-k_n(w_0)x - (w_0 - n\Omega)t]} \quad \text{and} \quad u^t(x,t) = \sum_{n=-\infty}^{\infty} u_n^t e^{i[k_n(w_0)(x-L) - (w_0 - n\Omega)t]} \quad \text{with}$$

$k_n(W_0) = \dfrac{W_0 - n\Omega}{W_0} k_0$ by applying the continuity conditions of pressure ($=B\dfrac{\partial u}{\partial x}$) and displacement on both surfaces of the slab, we finally get the transmissions, as shown in Figs. 3(a) and (b) by the black dots. These theoretical results for transmissions, giving the intensities of each monochromatic component at $w = w_0 + n\Omega$ ($n = 0, \pm 1, \pm 2, \pm 3\mathbf{L}$), normalized to the that of incident wave, $|u_n^t / u_0|^2$, capture the experimental features well.

It is well known that, for conventional spatial phononic crystals, the contrast in the properties of the comprising components, or the modulation-strength, plays a critical role in the properties of the phononic crystals. This is also true for the TPCs as shown below. To investigate experimentally the influence of modulation-strength on the transmission of the TPCs, we fill one chamber of the TPC with a mixture of $CO_2$ and air (instead of pure $CO_2$ used above), and the properties of the mixing gas, including the density and bulk modulus are tuned by controlling the relative concentration of $CO_2$. Transmission spectra measurements were performed in three cases, in which the relative concentration of $CO_2$ is selected to be 25%, 50% and 75%, respectively, which correspond to three increasing modulation-strengths. The results are shown in Fig. 5. Obviously, with the increasing of the modulation-strength, the split peaks have high intensities, and the frequency conversion is more remarkable.

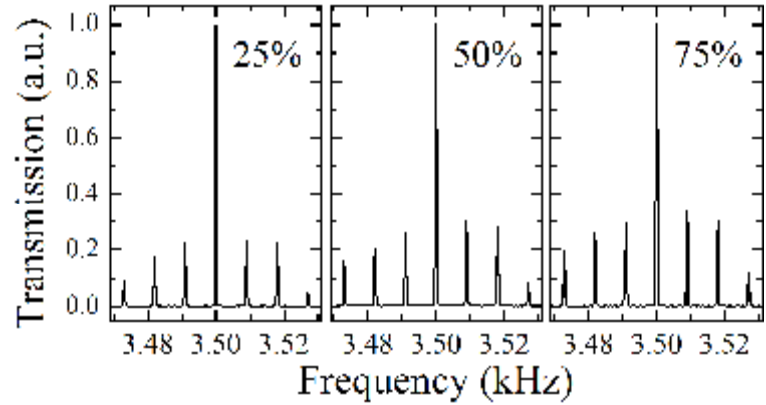

FIG. 5. Measured transmission spectra for the TPCs, with one chamber in the cylinder filled with the $CO_2$-air mixture and the $CO_2$ volume concentration: 25%, 50%, 75%, the incident wave frequency $w_0 = 3.5$ kHz.

In summary, we have given an instance of the temporal phononic crystals (TPCs) by rotationally swapping different gases between the sound source and detector. Prominent frequency conversions, including up-conversion and down-conversion are demonstrated experimentally in the transmission spectra, without involving any nonlinearity in system. These experimental results agree with the theoretical calculations based on the plane wave expansion in time domain. The TPCs may have potential applications in frequency converters, modulators, etc..

**Acknowledgement:**

This work is supported by the National Natural Science Foundation of China

(Grant Nos. 11174225, 11004155, 50425206); Open Foundation from State Key Laboratory of Applied Optics of China; and the Program for New Century Excellent Talents (NCET-11-0398).